\documentclass[11pt]{article}

\usepackage{amsmath}

\begin{document}

\title{
 V.V. Kisel,  E.M. Ovsiyuk,  \\  O.V. Veko, V.M. Red'kov. \\[3mm]
Quantum mechanics for a vector particle \\in the magnetic field on 4-dimensional sphere
 \\  [5mm] {\small Belarusian  State Pedagogical University \\
Mozyr State Pedagogical University, \\
Institute of Physics, NAS of Belarus\\
e.ovsiyuk@mail.ru , redkov@dragon.bas-net.by}}

\date{}

\maketitle

\begin{abstract}

Quantum-mechanical wave equation for a particle with spin 1 is investigated
in presence of external magnetic field in spaces with non-Euclidean geometry with constant
 positive curvature. Separation of the variable   is performed; differential equations in the variable
 $r$
  are solved in hypergeometric functions. The study of z-dependence of the wave function has been
  reduced to a system of three linked ordinary differential 2-nd order equations; till now the system in $z$
  variable is not solved.

\end{abstract}

\subsection*{1.  Introduction, setting the problem}

 In the present paper, we consider a quantum-mechanical problem  a particle with spin $1$ described
 by the Duffin--Kemmer in  3-dimensional Riemann space model
 in presence of the external magnetic field -- relevant publications see in [1--30].

 Initial  matrix wave equation of Duffin--Kemmer for a spin 1
 particle  has the for (we adhere notation \cite{Book-1})

 $$
\left \{ \beta^{c} [\; i \hbar \; (\;  e_{(c)}^{\beta}
\partial_{\beta}  + {1\over 2} J^{ab} \gamma_{abc}  \; )\;
-\; {e \over c} A_{c}  \; ] \; - \;  mc  \right \} \Psi  = 0 \; ,
\eqno(1.1)
$$

\noindent where   $\gamma_{abc}$ stand for Ricci rotation
coefficients
$$
\gamma_{bac} = - \gamma_{abc} = - e_{(b)\beta ; \alpha } \;
e_{(a)}^{\beta} e_{(c)}^{\alpha},
$$

\noindent $A_{a} = e_{(a)}^{\beta}  A_{\beta} $  are tetrad
components of an electromagnetic 4-vector  $A_{\beta} $; $J^{ab} =
(\beta ^{a}\beta ^{b}- \beta^{b} \beta ^{a})$ stand for generators of 10-dimensional
representation of the Lorentz group.
Below we will use shortened notation
б®Єа йҐ­Ёп $ e /c \hbar \Longrightarrow e, \; mc/ \hbar
\Longrightarrow M$.

In Olevsky paper  \cite{Olevsky} under the number  $XI$ the following coordinates are were specified
$$
dS^{2} = c^{2} dt^{2} - \rho^{2} \; [ \; \cos^{2} z ( d r^{2} +
\sin^{2} r \; d \phi^{2} ) + dz^{2}\; ]\; ,
$$
$$
z \in [-\pi /2 , + \pi /2 ]\; , \qquad r \in [0, + \pi ] , \qquad
\phi \in [0, 2 \pi ] \; .
\eqno(1.2)
$$

 Generalization of the concept of an uniform magnetic field for
 the curved model $S_{3}$
 is given by the following  potential
 $$
 A_{\phi} = -2B \sin^{2} {r \over 2} = B\; ( \cos r -1 )\; .
\eqno(1.3)
$$

\noindent To this potential there correspond a single non-vanishing component of the electromagnetic tensor
 $ F_{\phi r} =
\partial_{\phi}A_{r} - \partial_{r} A_{\phi} = B  \sin r $; this tensor satisfies Maxwell equations in $S_{3}$.

Let us consider eq.
(1.3) in the space  $S_{3}$. To cylindric coordinates $x^{\alpha}= (t,r,\phi,z)$
 there corresponds the tetrad
 $$
 e_{(a)}^{\beta}(x) = \left |
\begin{array}{llll}
1 & 0 & 0 & 0 \\
0 & \cos^{-1}z & 0 & 0 \\
0 & 0 & \cos^{-1}z\;\sin^{-1} r & 0 \\
0 & 0 & 0 & 1
\end{array} \right | \; .
\eqno(1.4)
$$

 Relevant Christoffel symbols  and Ricci rotation coefficients
 are
 $$
\Gamma^{r}_{\;\;jk } = \left | \begin{array}{ccc}
0 & 0 & -\mbox{tg}\;z \\
0 & - \sin r \cos r & 0 \\
- \mbox{tan}\;z & 0 & 0
\end{array} \right |  ,
$$
$$
\Gamma^{\phi}_{\;\;jk } = \left | \begin{array}{ccc}
0 & \mbox{cot}\; r & 0\\
\mbox{cot}\; r & 0 &- \mbox{tan}\; z \\
0 & -\mbox{tan}\; z & 0
\end{array} \right |  ,
$$
$$
\Gamma^{z}_{\;\;jk } = \left | \begin{array}{ccc}
\sin z \cos z & 0 & 0\\
0 & \sin z \; \cos z \sin^{2} r & 0 \\
0 & 0 & 0
\end{array} \right | \; ,
$$
$$
\gamma_{12 2} =
 { 1 \over \cos z \tan r} \; , \qquad
 \gamma_{31 1} =
 -\tan z\; , \qquad \gamma_{32 2} =
 -\tan z\; .
 \eqno(1.5)
 $$

So, general covariant Duffin--Kemmer equation   (1.1) takes the form

$$
\left \{  i \beta^{0} { \partial \over \partial t} + {1 \over \cos z} \left (  i  \beta^{1}
{ \partial \over \partial r}        + \beta^{2}      { i  \partial _{ \phi}- e B
(\cos r -1) + i J^{12} \cos r  \over \sin r}  \right )   \right.
$$
$$
 \left.
  +
   i\beta^{3}  { \partial \over \partial  z}  +  i   {\sin z \over \cos z}  \; ( \beta^{1} J^{13}    +   \beta^{2} J^{23} )
          - M   \right \} \Psi  = 0 \; ,
$$
$$
\eqno(1.6)
$$

In the limit of flat Minkowaki space, eq. (1.6) becomes simpler
$$
\left \{  i \beta^{0} {\partial \over \partial t} + i
  \beta^{1}    { \partial \over \partial r}        +
\beta^{2}      { i \partial _{ \phi} + e B  r^{2} /2  + i J^{12}
\over  r} +    i\beta^{3}  { \partial \over \partial   z} - M   \right \} \Psi  =
0 \; .
$$
$$
\eqno(1.7)
$$

To separate the variable we will need an explicit representation
for Duffin--Kemmer matrices
 $\beta^{a}$; most convenient for us is the  cyclic representation; in particular, then
 $J^{12}$ is diagonal
 (we will  use  blocks  structure in accordance with the structure
$1-3-3-3$) :
$$
\beta^{0} = \left | \begin{array}{rrrr}
 0       &   0        &  0  &  0 \\
 0  &  0       &  i  & 0  \\
  0  &   -i       &   0  & 0\\
   0  &  0       &   0  & 0
\end {array}
\right |, \qquad
\beta^{i} = \left |
\begin{array}{rrrrr}
  0       &  0       &    e_{i}  & 0       \\
    0   &  0       &   0      & \tau_{i} \\
   -e_{i}^{+}  &  0       &   0      & 0       \\
   0       &  -\tau_{i}&   0      & 0
\end {array} \right | \; ,
\eqno(1.8)
$$

\noindent where   $e_{i}, \; e_{i}^{\;t}, \; \tau_{i}$ designate
$$
e_{1} = {1 \over \sqrt{2}} ( -i, \; 0  , \; i )\; , \qquad e_{2} =
{1 \over \sqrt{2}} ( 1 , \; 0  , \;  1 )\; , \qquad e_{3} = ( 0 ,
i  , 0)\; , \;
$$
$$
\tau_{1} = {1 \over \sqrt{2}} \left |  \begin{array}{ccc} 0  &  1
&  0  \\ 1 &  0  &  1  \\ 0  &  1  &  0
\end{array} \right | ,
\tau_{2}= {1 \over \sqrt{2}} \left |
\begin{array}{ccc} 0  &  -i  &  0  \\ i & 0  &  -i  \\ 0  &  i  &
0
\end{array} \right | ,
  \tau_{3} =   \left |
\begin{array}{rrr} 1  &  0  &  0  \\ 0  &  0  &  0   \\ 0  &  0
&  -1
\end{array} \right |  =  s_{3}\; .
$$
$$
\eqno(1.9)
$$

\noindent Entering eq.  (1.6), the matrix  $J^{12}$ is
$$
 J^{12} =  \beta^{1} \beta^{2} -  \beta^{2} \beta^{1}
=
$$
$$
= \left | \begin{array}{cccc}
-e_{1}e_{2}^{+}  + e_{2}e_{1}^{+}& 0 & 0 & 0 \\
0 &-\tau_{1}\tau_{2} + \tau_{2} \tau_{1} & 0 & 0 \\
0 & 0 & -e_{1}^{+} \bullet e_{2} + e_{2}^{+} \bullet e_{1} & 0 \\
0 & 0 & 0 & - \tau_{1}\tau_{2} + \tau_{2} \tau_{1}
\end{array} \right | =
$$
$$
=  -i \left | \begin{array}{cccc}
0 & 0  &  0 & 0  \\
0 &   \tau_{3}  & 0 & 0 \\
0 & 0 &  \tau_{3} & 0 \\
0 & 0 & 0 &  \tau_{3}
\end{array} \right | = -iS_{3}.
\eqno(1.8)
$$

\subsection*{2. Separation of the variables
 }

Let us rewrite eq. (1.6) in the form
$$
\left [   i \beta^{0} \cos z  {\partial \over \partial t} + i   \beta^{1}
{\partial \over \partial r}        + \beta^{2}      { i \partial _{ \phi}- e B
(\cos r -1) + i J^{12} \cos r  \over  \sin r}   \right.
$$
$$
 \left.
  +
   i\beta^{3}  \cos z  {\partial \over \partial  z}  +  i \sin z ( \beta^{1} J^{13}    +
     \beta^{2} J^{23} )         - \cos z  M   \right ] \Psi  = 0 \; .
$$
$$
\eqno(2.1)
$$

To separate the variables, we will use the following substitution
for the wave function
$$
\Psi = e^{-i\epsilon t  }  e^{im\phi}   \left |
\begin{array}{c}
\Phi_{0}  (r,z) \\
\vec{\Phi}(r,z)  \\
\vec{E} (r,z)  \\
\vec{H} (r,z)
\end{array} \right | .
\eqno(2.2)
$$

\noindent Eq. (2.1)  leads us to  (let  $m + B (1 - \cos r) = \nu
(r)$)
$$
\left \{  \epsilon  \cos z  \;  \beta^{0}   +  i   \beta^{1}\;
{\partial \over  \partial r }        -
\beta^{2}    \;   { \nu (r)   -  \cos r \; S_{3}    \over  \sin r}   \right.
$$
$$
 \left.
  +
   i\beta^{3} \;  \cos z\;  {\partial \over \partial z}  +
     i  \; ( \beta^{1} J^{13}     +   \beta^{2} J^{23} )\;  \sin z       - \cos z \; M   \right \}
\left | \begin{array}{c}
\Phi_{0}  (r,z) \\
\vec{\Phi} (r,z) \\
\vec{E} (r,z) \\
\vec{H} (r,z)
\end{array} \right |  = 0 \; ,
\eqno(2.3)
$$

With the help of auxiliary relations

$$
 J^{13} =  \beta^{1} \beta^{3} -  \beta^{3} \beta^{1}
=
$$
$$
= \left | \begin{array}{cccc}
-e_{1}e_{3}^{+}  + e_{3}e_{1}^{+}& 0 & 0 & 0 \\
0 &-\tau_{1}\tau_{3} + \tau_{3} \tau_{1} & 0 & 0 \\
0 & 0 & -e_{1}^{+} \bullet e_{3} + e_{3}^{+} \bullet e_{1} & 0 \\
0 & 0 & 0 & - \tau_{1}\tau_{3} + \tau_{3} \tau_{1}
\end{array} \right | =
$$
$$
=  i \left | \begin{array}{cccc}
0 & 0  &  0 & 0  \\
0 &   \tau_{2}  & 0 & 0 \\
0 & 0 &  \tau_{2} & 0 \\
0 & 0 & 0 &  \tau_{2}
\end{array} \right | = iS_{2}\;,
$$
$$
 J^{23} =  \beta^{2} \beta^{3} -  \beta^{3} \beta^{2}
=
$$
$$
= \left | \begin{array}{cccc}
-e_{2}e_{3}^{+}  + e_{3}e_{2}^{+}& 0 & 0 & 0 \\
0 &-\tau_{2}\tau_{3} + \tau_{3} \tau_{2} & 0 & 0 \\
0 & 0 & -e_{2}^{+} \bullet e_{3} + e_{3}^{+} \bullet e_{2} & 0 \\
0 & 0 & 0 & - \tau_{2}\tau_{3} + \tau_{3} \tau_{2}
\end{array} \right | =
$$
$$
=  -i \left | \begin{array}{cccc}
0 & 0  &  0 & 0  \\
0 &   \tau_{1}  & 0 & 0 \\
0 & 0 &  \tau_{1} & 0 \\
0 & 0 & 0 &  \tau_{1}
\end{array} \right | = -iS_{1}\; ,
$$

\noindent
we get
$$
( \beta^{1} J^{13}     + \beta^{2} J^{23} ) =
$$
$$
= i \left |
\begin{array}{rrrrr}
  0       &  0       &    e_{1}  & 0       \\
    0   &  0       &   0      & \tau_{1} \\
   -e_{1}^{+}  &  0       &   0      & 0       \\
   0       &  -\tau_{1}&   0      & 0
\end {array} \right |    \left | \begin{array}{cccc}
0 & 0  &  0 & 0  \\
0 &   \tau_{2}  & 0 & 0 \\
0 & 0 &  \tau_{2} & 0 \\
0 & 0 & 0 &  \tau_{2}
\end{array} \right | -
$$
$$
-
i \left | \begin{array}{rrrrr}
  0       &  0       &    e_{2}  & 0       \\
    0   &  0       &   0      & \tau_{2} \\
   -e_{2}^{+}  &  0       &   0      & 0       \\
   0       &  -\tau_{2}&   0      & 0
\end {array} \right |  \left | \begin{array}{cccc}
0 & 0  &  0 & 0  \\
0 &   \tau_{1}  & 0 & 0 \\
0 & 0 &  \tau_{1} & 0 \\
0 & 0 & 0 &  \tau_{1}
\end{array} \right |
$$
$$
=i \; \left | \begin{array}{cccc}
0 & 0 & e_{1} \tau_{2} - e_{2}\tau_{1} & 0 \\
0 & 0 & 0 & \tau_{1} \tau_{2} -  \tau_{2} \tau_{1 } \\
0 & 0 &  0   & 0 \\
0 & -\tau_{1} \tau_{2} +  \tau_{2} \tau_{1} & 0 & 0
\end{array} \right | =
 \left | \begin{array}{cccc}
0 & 0 &  -2 e_{3}  & 0 \\
0 & 0 & 0 &  - \tau_{3}  \\
0 & 0 &  0   & 0 \\
0 & + \tau_{3} & 0 & 0
\end{array} \right |
$$

\noindent eq. (2.3)   can be presented as
$$
 \left  [ \;  \epsilon \; \cos z\;   \left | \begin{array}{rrrr}
 0       &   0        &  0  &  0 \\
 0  &  0       &  i  & 0  \\
  0  &   -i       &   0  & 0\\
   0  &  0       &   0  & 0
\end{array} \right |    + i\;  \left |
\begin{array}{rrrrr}
  0       &  0       &    e_{1}  & 0       \\
    0   &  0       &   0      & \tau_{1} \\
   -e_{1}^{+}  &  0       &   0      & 0       \\
   0       &  -\tau_{1}&   0      & 0
\end {array} \right | {\partial \over \partial r }  \right.
$$
$$
 - {1 \over  \sin r  } \; \left |
\begin{array}{rrrrr}
  0       &  0       &    e_{2}  & 0       \\
    0   &  0       &   0      & \tau_{2} \\
   -e_{2}^{+}  &  0       &   0      & 0       \\
   0       &  -\tau_{2}&   0      & 0
\end {array} \right |
 (   \nu     - \cos  r \; S_{3})
 $$
 $$
   + i \cos z\;   \left |
\begin{array}{rrrrr}
  0       &  0       &    e_{3}  & 0       \\
    0   &  0       &   0      & \tau_{3} \\
   -e_{3}^{+}  &  0       &   0      & 0       \\
   0       &  -\tau_{3}&   0      & 0
\end {array} \right | {\partial \over \partial z}
$$
$$
\left.
+ i \sin z \left | \begin{array}{cccc}
0 & 0 &  -2 e_{3}  & 0 \\
0 & 0 & 0 &  - \tau_{3}  \\
0 & 0 &  0   & 0 \\
0 & + \tau_{3} & 0 & 0
\end{array} \right |
  - M\; \cos z \;  \right  ]  \left | \begin{array}{c}
\Phi_{0} \\
\vec{\Phi} \\
\vec{E} \\
\vec{H}
\end{array} \right |
 = 0 \;  .
$$
$$
\eqno(2.4)
$$

In block form it is written
$$
   i e_{1}  \partial _{r } \vec{E}
   -  {1 \over  \sin r} \; e_{2} (\nu  - \cos r \,s_{3} )  \vec{E}
    + i  (\cos z\;  \;  \partial _{z }  - 2 \sin z ) e_{3}\vec{E}= M \cos z   \; \Phi_{0} \; ,
$$
$$
i \epsilon \cos z\, \vec{E} + i \tau_{1} \partial_{r} \vec{H} -
{\tau_{2} \over \sin r } (   \nu  - \cos r \, s_{3} ) \vec{H}  +
i  ( \cos z\; \partial _{z }   - \sin z ) \tau_{3}\vec{H} =M \cos z   \vec{\Phi}  ,
$$
$$
-i\epsilon \cos z\, \vec{\Phi}  -i e_{1}^{+} \partial_{r}  \Phi_{0} +
{\nu \over  \sin r} \,e_{2}^{+} \Phi_{0}  - i \cos z\; e_{3}^{+}\partial _{z }
\Phi_{0} = M \cos z \;  \vec{E}\; ,
$$
$$
-i \tau_{1} \partial_{r} \vec{\Phi} + { (\nu -\cos r \, s_{3})
\over  \sin r} \,\tau_{2}  \vec{\Phi}  - i ( \cos z\; \partial _{z } - i\sin z ) \tau_{3}\vec{\Phi}= M \cos z
\; \vec{H}\; .
$$
$$
\eqno(2.5)
$$

After simple calculation,  we arrive
at a  system of 10 equations (let $\gamma =1 / \sqrt{2}$)

$$
  \gamma  (  { \partial  E_{1} \over \partial  r} - {\partial E_{3} \over \partial r
})  -  {\gamma \over  \sin r }   \left [  (\nu -\cos r ) E_{1} +
(\nu +\cos r) E_{3}   \right ]   -
$$ $$ - (  \cos z {\partial
\over
\partial z} - 2 \sin z ) E_{2}
 = M \cos z   \Phi_{0} \; ,
$$
$$
+i \epsilon \cos z  E_{1}    +   i \gamma  {\partial H_{2} \over \partial  r}
 +   i\gamma     { \nu \over  \sin r }   H_{2}
 + i  ( \cos z   {\partial   \over \partial  z}- \sin z ) H_{1}= M \cos z \Phi_{1}\;,
$$
$$
+i \epsilon  \cos z  E_{2}  +  i \gamma ( {\partial  H_{1} \over \partial  r
} + {\partial  H_{3} \over \partial  r } ) -
 {i\gamma
\over  \sin r }  [  (\nu -\cos r)  H_{1}  -
$$
$$
-  (\nu + \cos r)  H_{3}   ]  =
M \cos z  \Phi_{2}\;,
$$
$$
+i \epsilon  \cos z  E_{3}    +  i \gamma\; {\partial   H_{2} \over \partial
r}  -   i \gamma  {\nu  \over  \sin r }  H_{2} - i  (\cos z
{\partial  \over \partial z} - \sin z ) H_{3} =
M \cos z   \Phi_{3}
$$
$$
\eqno(2.6)
$$

$$
-i  \epsilon  \cos z \Phi_{1}  +  \gamma   {\partial
\Phi_{0} \over d r}   +
 \gamma  {\nu \over  \sin r }   \Phi_{0} = M \cos z   E_{1}\;,
$$
$$
-i \epsilon  \cos z \Phi_{2}  - \cos z  {\partial  \Phi_{0} \over \partial  z} = M \cos z
E_{2}\;,
$$
$$
-i \epsilon \cos z \Phi_{3}  -   \gamma   {\partial  \Phi_{0} \over \partial
r}  +
 \gamma {\nu   \over  \sin r }   \Phi_{0} = M \cos z  E_{3}\;,
\eqno(2.7)
$$

$$
-i  \gamma   {\partial  \Phi_{2} \over \partial  r}  -  i \gamma   {\nu   \over  \sin r }  \Phi_{2}-
i ( \cos z{\partial  \over \partial  z} - \sin z ) \Phi_{1} = M \cos z  H_{1}\;,
$$
$$
 - i  \gamma  (  {\partial  \Phi_{1} \over \partial r } + { \partial \Phi_{3} \over \partial r })
+  {i \gamma \over  \sin r }  [ (\nu -\cos r) \Phi_{1} -
(\nu+\cos r)\Phi_{3}  ]  = M  \cos z  H_{2} \; ,
$$
$$
-i  \gamma   {\partial  \Phi_{2} \over \partial r}  +
 i \gamma {\nu \over  \sin r }   \Phi_{2}
  +i (  \cos z  { \partial   \over \partial  z}-  \sin z ) \Phi_{3}=  M \cos z   H_{3}\;.
$$
$$
\eqno(2.8)
$$

With the help of substitutions
$$
H_{1} = {h_{1} \over \cos z} \; , \qquad
( \cos z \; { \partial  \over \partial z}- \sin z ) H_{1} = {\partial   h_{1} \over \partial z}\; ,
$$
$$
H_{3} = {h_{3} \over \cos z} \; , \qquad
( \cos z \; { \partial   \over \partial z}- \sin z ) H_{3} = {\partial  h_{3}  \over \partial z}\; ,
$$
$$
\Phi_{1} = {\varphi_{1} \over \cos z} \; , \qquad
( \cos z \; { \partial  \over \partial  z}- \sin z ) \Phi_{1} = {\partial  \varphi_{1} \over \partial z} \; ,
$$
$$
\Phi_{3} = {\varphi_{3} \over \cos z} \; , \qquad
( \cos z \; {\partial  \over \partial  z}- \sin z ) \Phi_{3} = {\partial  \varphi_{3}\over \partial z} \; ,
$$
$$
E_{2} = {e_{2}  \over  \cos ^{2}z }\; , \qquad
(  \cos z \;{\partial  \over \partial z } -2 \sin z ) E_{2} = {1 \over \cos z }  {\partial  e_{2} \over \partial z } \; ,
$$
$$
E_{1} = { e_{1} \over  \cos z}\;, \qquad E_{3} = { e_{3} \over  \cos z}\;,
$$
$$
\Phi_{0} =  {\varphi_{0} \over  \cos^{2} z} \; ,\qquad
\Phi_{2} =  {\varphi_{2} \over  \cos^{2} z} \; , \qquad
H_{2} =  {h_{2} \over  \cos^{2} z}\; ,
$$
$$
\eqno(2.9)
$$

\noindent we get a more simple system

$$
  \gamma \; (  {\partial e_{1} \over \partial r} - {\partial  e_{3} \over \partial r
})  -  {\gamma \over  \sin r }   \left [  (\nu -\cos r )
e_{1} + (\nu +\cos r)e_{3}   \right ]   -   {\partial   e_{2} \over \partial z }
 = M   \varphi_{0} \; ,
$$
$$
+i \epsilon    e_{1}    +   i {\gamma  \over \cos^{2} z}  ( {\partial   \over \partial  r}
 +      { \nu \over  \sin r })\;    h_{2}
 + i  {\partial    h_{1} \over \partial z} = M    \varphi _{1}\;,
$$
$$
+i \epsilon    e_{2}   +  i \gamma   ( {\partial  h_{1} \over \partial  r
} + {\partial  h_{3} \over \partial  r } ) -
 {i\gamma
\over  \sin r  } \left [  (\nu -\cos r)  h_{1}  -   (\nu +\cos r)  h_{3}  \right ]  =
M   \varphi_{2}\;,
$$
$$
+i \epsilon    e_{3}    +  i {\gamma \over \cos^{2} z}  ( {\partial    \over \partial
r}  -    {\nu  \over  \sin r } )  h_{2} - i  {\partial    h_{3} \over \partial z}=
M    \varphi_{3} \; .
$$
$$
\eqno(2.10)
$$

$$
-i  \epsilon   \varphi_{1}   +   {\gamma  \over  \cos^{2} z}  (  {\partial
 \over \partial  r}  +
  {\nu \over  \sin r }  )  \varphi_{0} = M     e_{1}\;,
$$
$$
-i \epsilon   \varphi_{2}   - ( {\partial  \over \partial z}  + 2 {\sin z \over \cos z})  \varphi _{0} =
M   e_{2} \;,
$$
$$
\\
-i \epsilon  \varphi _{3}  -  {\gamma \over \cos^{2} z}   (  {\partial   \over \partial
r}  -  {\nu   \over  \sin r } )\; \varphi_{0} = M    e_{3}\;,
$$
$$
\eqno(2.11)
$$

$$
-i  {\gamma \over \cos ^{2} z}  \; ( { \partial   \over d r}  +    {\nu   \over  \sin r } )  \varphi_{2} \;-
i  {\partial   \varphi_{1} \over \partial z} = M h_{1}\;,
$$
$$
 - i   \gamma   (  { \partial  \varphi_{1} \over d r } + {d \varphi_{3} \over d r })\;
+  {i \gamma \over  \sin r }  [  (\nu -\cos r) \varphi_{1} -
(\nu+\cos r)\varphi_{3}   ]  = M    h_{2} \; ,
$$
$$
-i  {\gamma  \over \cos^{2} z}  (  {\partial   \over \partial  r}  - {\nu \over  \sin r })  \varphi_{2}
  +i  {\partial   \varphi_{3} \over \partial z} =  M     h_{3}\;.
$$
$$
\eqno(2.12)
$$

\noindent These equation can be transformed to the form

$$
  \gamma \;  ( {\partial  \over \partial r} -  { \nu -\cos r  \over  \sin r } )  \; e_{1}
  - \gamma  ( {\partial   \over \partial r
}  +  { \nu +\cos r \over  \sin r } ) \;  e_{3}
-   {\partial e_{2} \over
\partial z }
 = M   \; \varphi_{0} \; ,
$$
$$
    i \gamma   (  {\partial   \over \partial  r }  - { \nu -\cos r \over  \sin r  }   ) \;  h_{1}
  +  i \gamma  ( {\partial  \over \partial  r }
   + { \nu +\cos r  \over  \sin r  }   )\; h_{3}
    + i \epsilon  \;   e_{2}   = M  \;  \varphi_{2}\;,
    $$
    $$
    {i\gamma  \over \cos^{2} z}  ({\partial   \over \partial  r}
 +      { \nu \over  \sin r })\;    h_{2} +i \epsilon  \;   e_{1}
 + i  {\partial    h_{1} \over \partial z} = M  \;   \varphi _{1}\;,
$$
$$
   {i\gamma \over \cos^{2} z}  (
{\partial    \over \partial r}  -    {\nu  \over  \sin r } ) h_{2} +i \epsilon  \;   e_{3}
- i  {\partial    h_{3} \over \partial z}= M    \; \varphi_{3}\; ,
$$
$$
\eqno(2.13a)
$$

$$
   {\gamma  \over  \cos^{2} z} \;  ( {\partial  \over \partial  r}  +
  {\nu \over  \sin r }  )  \; \varphi_{0}  - i  \epsilon  \;  \varphi_{1} = M \;     e_{1}\;,
$$
$$
-  {i\gamma \over \cos ^{2} z}  \; ( { \partial   \over d r}  +
{\nu   \over  \sin r } )  \varphi_{2} \;- i  {\partial
\varphi_{1} \over \partial z} = M  \; h_{1}\;,
$$
$$
 -  {\gamma \over \cos^{2} z}   (
{\partial   \over \partial r}  -  {\nu   \over  \sin r } )\;
\varphi_{0}  -i \epsilon  \;\varphi _{3}  = M    \; e_{3}\;,
$$
$$
-  {i\gamma  \over \cos^{2} z}  (  {\partial   \over \partial  r} - {\nu \over  \sin r }) \;  \varphi_{2}
  + i  {\partial   \varphi_{3} \over \partial z} =  M  \;    h_{3}\;,
$$
$$
-i \epsilon   \varphi_{2}   - ( {\partial  \over \partial z}  + 2
{\sin z \over \cos z})  \varphi _{0} = M  \;  e_{2} \;,
$$
$$
 - i   \gamma  (   { \partial   \over \partial r  } -  { \nu -\cos r \over  \sin r }    ) \; \varphi_{1}
 - i   \gamma  ( {\partial \over \partial r } \;  + { \nu+\cos r  \over  \sin r } ) \; \varphi_{3}     = M  \;   h_{2} \; .
$$
$$
\eqno(2.13b)
$$

Let us introduce a shortened notation

$$
\gamma   ( {\partial  \over \partial r} +  { \nu -\cos r  \over  \sin r } ) = \hat{a}_{-} ,
\gamma   ( {\partial  \over \partial r} +  { \nu +\cos r  \over  \sin r } ) = \hat{a}_{+},
\gamma   ( {\partial  \over \partial r} +  { \nu   \over  \sin r } ) = \hat{a}  ,
$$
$$
\gamma   (- {\partial  \over \partial r} +  { \nu -\cos r  \over  \sin r } ) = \hat{b}_{-} ,
\gamma   (- {\partial  \over \partial r} +  { \nu +\cos r  \over  \sin r } ) = \hat{b}_{+},
\gamma   (- {\partial  \over \partial r} +  { \nu   \over  \sin r } ) = \hat{b} \; ,
$$
$$
\eqno(2.14)
$$

\noindent
then the above equations read

$$
  -\hat{b}_{-}  \; e_{1}   - \hat{a}_{+}  \;  e_{3}
-   {\partial e_{2} \over
\partial z }
 = M   \; \varphi_{0} \; ,
$$
$$
   - i  \hat{b}_{-}  \;  h_{1}    +  i \hat{a}_{+} \; h_{3}
    + i \epsilon  \;   e_{2}   = M  \;  \varphi_{2}\;,
    $$
    $$
        {i \over \cos^{2} z}  \hat{a}  \;    h_{2} +i \epsilon  \;   e_{1}
 + i  {\partial    h_{1} \over \partial z} = M  \;   \varphi _{1}\;,
$$
$$
  - {i \over \cos^{2} z}  \hat{b}\;  h_{2} +i \epsilon  \;   e_{3}
- i  {\partial    h_{3} \over \partial z}= M    \; \varphi_{3}\; ,
\eqno(2.15)
$$
$$
   {1  \over  \cos^{2} z} \;  \hat{a} \; \varphi_{0}  - i  \epsilon  \;  \varphi_{1} = M \;     e_{1}\;,
$$
$$
-  {i \over \cos ^{2} z}  \; \hat{a}\;  \varphi_{2} \;- i  {\partial
\varphi_{1} \over \partial z} = M  \; h_{1}\;,
$$
$$
   {1 \over \cos^{2} z}  \;  \hat{b} \;
\varphi_{0}  -i \epsilon  \;\varphi _{3}  = M    \; e_{3}\;,
$$
$$
  {i  \over \cos^{2} z}  \; \hat{b} \;  \varphi_{2}
  + i  {\partial   \varphi_{3} \over \partial z} =  M  \;    h_{3}\;.
$$
$$
-i \epsilon   \varphi_{2}   - ( {\partial  \over \partial z}  + 2
{\sin z \over \cos z})  \varphi _{0} = M  \;  e_{2} \;,
$$
$$
  i   \hat{b}_{-}  \; \varphi_{1}
 - i  \hat{a}_{+} \; \varphi_{3}     = M  \;   h_{2} \; ,
\eqno(2.16)
$$

We can note that turning back to  $\Phi_{0}$, we get a simple system as well

$$
  -\hat{b}_{-}  \; e_{1}   - \hat{a}_{+}  \;  e_{3}
-   {\partial e_{2} \over
\partial z }
 = M   \; \cos^{2} z \Phi_{0} \; ,
$$
$$
   - i  \hat{b}_{-}  \;  h_{1}    +  i \hat{a}_{+} \; h_{3}
    + i \epsilon  \;   e_{2}   = M  \;  \varphi_{2}\;,
    $$
    $$
        {i \over \cos^{2} z}  \hat{a}  \;    h_{2} + i \epsilon  \;   e_{1}
 + i  {\partial    h_{1} \over \partial z} = M  \;   \varphi _{1}\;,
$$
$$
  - {i \over \cos^{2} z}  \hat{b}\;  h_{2} + i \epsilon  \;   e_{3}
- i  {\partial    h_{3} \over \partial z}= M    \; \varphi_{3}\; ,
\eqno(2.17)
$$

$$
     \hat{a} \; \Phi_{0}  - i  \epsilon  \;  \varphi_{1} = M \;     e_{1}\;,
$$
$$
-  {i \over \cos ^{2} z}  \; \hat{a} \;  \varphi_{2} \;- i  {\partial
\varphi_{1} \over \partial z} = M  \; h_{1}\;,
$$
$$
    \hat{b} \;\Phi_{0}  -i \epsilon  \;\varphi _{3}  = M    \; e_{3}\;,
$$
$$
  {i  \over \cos^{2} z}  \; \hat{b} \;  \varphi_{2}
  + i  {\partial   \varphi_{3} \over \partial z} =  M  \;    h_{3}\;.
$$
$$
-i \epsilon \;   \varphi_{2}   - \cos^{2} z\;  {\partial  \Phi _{0}  \over \partial z}   = M  \;  e_{2} \;,
$$
$$
  i   \hat{b}_{-}  \; \varphi_{1}  - i  \hat{a}_{+} \; \varphi_{3}     = M  \;   h_{2} \; .
\eqno(2.18)
$$

Below we will work with equations  (2.17) --  (2.18).

\subsection*{3.  Transition to a  non-relativistic approximation}

\hspace{5mm} Excluding  from (2.17)--(2.18) non-dynamical variables
$\Phi_{0}, h_{1}, h_{2}, h_{3}$:

$$
{1 \over \cos^{2} z }
 (-\hat{b}_{-}  \; e_{1}   - \hat{a}_{+}  \;  e_{3} -   {\partial
e_{2} \over \partial z })     = M   \;  \Phi_{0} \; ,
$$
$$
-  {i \over \cos ^{2} z}  \; \hat{a} \;  \varphi_{2} \;- i
{\partial \varphi_{1} \over \partial z} = M  \; h_{1}\;,
$$

$$
 i   \hat{b}_{-}  \; \varphi_{1}  - i  \hat{a}_{+} \; \varphi_{3}     = M  \;   h_{2} \;
 ,$$

 $$
  {i  \over \cos^{2} z}  \; \hat{b} \;  \varphi_{2}
  + i  {\partial   \varphi_{3} \over \partial z} =  M  \;    h_{3}\;.
  \eqno(3.1)
  $$

\noindent we obtain 6 equations (grouping them in pair)

    $$
        {i \over \cos^{2} z}  \hat{a}  \;
        ( i   \hat{b}_{-}  \; \varphi_{1}  - i  \hat{a}_{+} \; \varphi_{3})
         + i \epsilon  \;   Me_{1}
 + i  {\partial     \over \partial z} (-  {i \over \cos ^{2} z}  \; \hat{a} \;  \varphi_{2} \;- i
{\partial \varphi_{1} \over \partial z})
 = M^{2}  \;   \varphi _{1}\;,
$$
$$
     \hat{a} \;
     {1 \over \cos^{2} z }
 (-\hat{b}_{-}  \; e_{1}   - \hat{a}_{+}  \;  e_{3} -   {\partial
e_{2} \over \partial z })
  - i  \epsilon  \;  M\varphi_{1} = M ^{2}\;     e_{1}\;,
$$
$$
\eqno(3.2a)
$$

$$
   - i  \hat{b}_{-}  \; ( -  {i \over \cos ^{2} z}  \; \hat{a} \;  \varphi_{2} \;-
   i {\partial \varphi_{1} \over \partial z})    +  i \hat{a}_{+} \;
({i  \over \cos^{2} z}  \; \hat{b} \;  \varphi_{2}
  + i  {\partial   \varphi_{3} \over \partial z} )
    + i \epsilon  \;   Me_{2}   = M ^{2} \;  \varphi_{2}\;,
    $$
$$
-i \epsilon \;   M \varphi_{2}   - \cos^{2} z\;  {\partial    \over \partial z}
{1 \over \cos^{2} z }
 (-\hat{b}_{-}  \; e_{1}   - \hat{a}_{+}  \;  e_{3} -   {\partial
e_{2} \over \partial z })    = M^{2}  \;  e_{2} \;,
$$
$$
\eqno(3.2b)
$$

$$
  - {i \over \cos^{2} z}  \hat{b}\;
  ( i   \hat{b}_{-}  \; \varphi_{1}  - i  \hat{a}_{+} \; \varphi_{3})
+ i \epsilon  \;   M e_{3}
- i  {\partial    \over \partial z}
({i  \over \cos^{2} z}  \; \hat{b} \;  \varphi_{2}
  + i  {\partial   \varphi_{3} \over \partial z})
  = M^{2}    \; \varphi_{3}\; ,
$$
$$
    \hat{b} \;
    {1 \over \cos^{2} z }
 (-\hat{b}_{-}  \; e_{1}   - \hat{a}_{+}  \;  e_{3} -   {\partial
e_{2} \over \partial z })    -i \epsilon  \; M \varphi _{3}  = M  ^{2}  \;
e_{3}\; .
$$
$$
\eqno(3.2c)
$$

Now we should introduce big $\Psi_{i}$ and small  $\psi_{i}$ components

$$
\varphi_{1} = \Psi_{1} + \psi_{1}\; , \qquad i e_{1} =  \Psi_{1} - \psi_{1}\; ,
$$
$$
\varphi_{2} = \Psi_{2} + \psi_{2}\; , \qquad i e_{2} =  \Psi_{2} - \psi_{2}\; ,
$$
$$
\varphi_{3} = \Psi_{3} + \psi_{3}\; , \qquad i e_{3} =  \Psi_{3} - \psi_{3}\; ,
$$

\noindent and in the same time separate the rest energy  by formal change
$\epsilon \Longrightarrow (\epsilon + M $) -- so we arrive at

$$
-{\hat{a} \hat{b}_{-} \over  \cos^{2} z} (\Psi_{1} + \psi_{1}) +
{\hat{a} \hat{a}_{+} \over  \cos^{2} z} (\Psi_{3} + \psi_{3})
 +  {\hat{a} \over \cos^{2} z} ({\partial \over  \partial z} + {2 \sin z \over  \cos z})    (\Psi_{2}+\psi_{2})
$$
$$
+  {\partial^{2} \over  \partial z^{2}}
(\Psi_{1} +\psi_{1})  + (\epsilon + M)
M  (\Psi_{1} - \psi_{1}) = M^{2} (\Psi_{1} +\psi_{1} ) \; ,
$$
$$
- {\hat{a} \hat{b}_{-} \over  \cos^{2} z} (\Psi_{1} - \psi_{1})
-
{\hat{a} \hat{a}_{+} \over  \cos^{2} z} (\Psi_{3} - \psi_{3})
-   {\hat{a} \over \cos^{2} z }   {\partial \over  \partial z  } (\Psi_{2} -\psi_{2})
$$
$$
+ (\epsilon +M)  M  (\Psi_{1} +\psi_{1}) =  M^{2} (\Psi_{1} - \psi_{1} ) \; ;
$$
$$
\eqno(3.3a)
$$
$$
-{\hat{b}_{-} \hat{a}  + \hat{a}_{+} \hat{b}  \over  \cos^{2} z} (\Psi_{2} + \psi_{2})
- \hat{b}_{-} {\partial \over  \partial z} (\Psi_{1} + \psi_{1})
- \hat{a}_{+} {\partial \over  \partial z} (\Psi_{3} + \psi_{3})
$$
$$
+  (\epsilon+M) M (\Psi_{2} - \psi_{2})  = M^{2} (\Psi_{2} +\psi_{2} ) \; ,
$$
$$
  \hat {b}_{-} \cos^{2} z\;  {\partial \over \partial z} {1 \over \cos^{2} z }
(\Psi_{1} - \psi_{1})
+ \hat {a}_{+} \cos^{2} z\;  {\partial \over \partial z} {1 \over \cos^{2} z }
(\Psi_{3} - \psi_{3})
$$
$$
+ \cos^{2} z\;  {\partial \over \partial z} {1 \over \cos^{2} z }{\partial \over  \partial z}
(\Psi_{2} - \psi_{2})  + (\epsilon+M)  M (\Psi_{2} + \psi_{2}) =  M^{2}  (\Psi_{2} - \psi_{2}) \; ;
$$
$$
\eqno(3.3b)
$$
$$
{\hat{b} \hat{b}_{-} \over  \cos^{2} z} (\Psi_{1} + \psi_{1}) -
{\hat{b} \hat{a}_{+} \over  \cos^{2} z} (\Psi_{3} + \psi_{3})   +
{1 \over \cos^{2} z} ( {\partial \over  \partial z} + {2 \sin z \over \cos z})   \hat{b} (\Psi_{2}+\psi_{2})+
$$
$$
+  {\partial^{2} \over  \partial z^{2}}
(\Psi_{3} +\psi_{3})  +(\epsilon+M)
M  (\Psi_{3} - \psi_{3}) = M^{2} (\Psi_{3} +\psi_{3} ) \; ,
$$
$$
 - {\hat{b} \hat{b}_{-} \over  \cos^{2} z} (\Psi_{1} - \psi_{1}) -
{\hat{b} \hat{a}_{+} \over  \cos^{2} z} (\Psi_{3} - \psi_{3}) -   {\hat{b}
\over \cos^{2} z }   {\partial \over  \partial z  } (\Psi_{2} -\psi_{2})
$$
$$
+ (\epsilon +M) M
(\Psi_{3} +\psi_{3}) =  M^{2} (\Psi_{3} - \psi_{3} ) \; .
$$
$$
\eqno(3.3c)
$$

Summing equation for each pair
and neglecting small components $\psi_{k}$  in comparison with big ones
 $\Psi_{k}$, we get

$$
\left ( -{ 2\over  \cos^{2} z} \; \hat{a} \hat{b}_{-}   +
 {\partial^{2} \over  \partial z^{2}}    + 2\epsilon M  \right ) \Psi_{1}  +
    2{ \sin z \over  \cos^{3} z}   \; \hat{a}  \Psi_{2}   = 0 \; ,
$$
$$
\left ( - { 2 \over   \cos^{2} z }  \;   \hat{b} \hat{a}_{+}
+
 {\partial^{2} \over  \partial z^{2}}
  +2\epsilon M \right )  \Psi_{3}   +2{ \sin z \over \cos^{3} z} \;  \hat{b} \Psi_{2} = 0 \; ,
$$
$$
\left ( -  {1  \over  \cos^{2} z} (\hat{b}_{-} \hat{a}  + \hat{a}_{+} \hat{b} )
+2\epsilon M   + {\partial^{2} \over  \partial z ^{2}} +2{ \sin z \over \cos z}{\partial \over  \partial z}
  \right )  \Psi_{2}
  $$
  $$
     +2 {\sin  z \over \cos z} (\hat{b}_{-} \Psi_{1} + \hat{a}_{+} \Psi_{3}) =0 \; .
\eqno(3.4a)
$$

It is a needed system in Pauli approximation.
In particular, for the case of flat space model we get much more simple system of three separated equations

$$
\left ( -  2 \hat{a} \hat{b}_{-}   +
 {\partial^{2} \over  \partial z^{2}}    + 2\epsilon M  \right ) \Psi_{1}   = 0 \; ,
$$
$$
\left ( -  2    \hat{b} \hat{a}_{+}
+
 {\partial^{2} \over  \partial z^{2}}
  +2\epsilon M \right )  \Psi_{3}    = 0 \; ,
$$
$$
\left ( -  (\hat{b}_{-} \hat{a}  + \hat{a}_{+} \hat{b} )
+2\epsilon M   + {\partial^{2} \over  \partial z ^{2}}
  \right ) \; \Psi_{2}  =0 \; ,
$$

\noindent where in definitions for $
\hat{a},  \hat{b}, \hat{a}_{-}, \hat{b}_{-}, \hat{a}_{+}, \hat{b}_{+}
$
some simplifications are to be performed
-- see   (2.14).

Equations (3.4a) can be transformed to a more symmetrical form if one make a substitution

$$
\Psi_{2} = \cos z \bar{\Psi}_{2} \;,
$$
$$
( {\partial^{2} \over  \partial z ^{2}} +2{ \sin z \over \cos z}{\partial \over  \partial z} ) \cos z \bar{\Psi}_{2} =
\cos z ( {\partial^{2} \over  \partial z ^{2}}  - {2 \over \cos^{2} z}  +1 ) \bar{\Psi}_{2} \; ,
\eqno(3.4b)
$$

Then, eqs. (3.4a) read

$$
\left ( -{ 2 \hat{a} \hat{b}_{-} \over  \cos^{2} z} \;    +
 {\partial^{2} \over  \partial z^{2}}    + 2\epsilon M  \right ) \Psi_{1}  +
    2{ \sin z \over  \cos^{2} z}   \; \hat{a}  \bar{\Psi}_{2}   = 0 \; ,
$$
$$
\left ( - { 2 \hat{b} \hat{a}_{+} \over   \cos^{2} z }  \;
+
 {\partial^{2} \over  \partial z^{2}}
  +2\epsilon M \right )  \Psi_{3}   +2{ \sin z \over \cos^{2} z} \;  \hat{b} \bar{\Psi}_{2} = 0 \; ,
$$
$$
\left ( -  { (\hat{b}_{-} \hat{a}  + \hat{a}_{+} \hat{b} +2 )  \over  \cos^{2} z}
  + {\partial^{2} \over  \partial z ^{2}} +2\epsilon M  +1 \right ) \bar{\Psi}_{2}
  $$
  $$
 +2 {\sin  z \over \cos^{2} z}  (\hat{b}_{-} \Psi_{1} + \hat{a}_{+} \Psi_{3}) =0 \; .
\eqno(3.4c)
$$

Let us introduce new functions

$$
\hat{b}_{-}   \Psi_{1} = G_{1} \; , \qquad \bar{\Psi}_{2} = G_{2}\; ,
\qquad \hat{a}_{+} \Psi_{3} = G_{3} \; ,
\eqno(3.5a)
$$

\noindent
eqs.  (3.4a) will give

$$
\left ( -{ 2\over  \cos^{2} z} \;  \hat{b}_{-} \hat{a}    +
 {\partial^{2} \over  \partial z^{2}}    + 2\epsilon M  \right ) G_{1}  +
    2{ \sin z \over  \cos^{2} z}   \; \hat{b}_{-} \hat{a} \; G_{2}   = 0 \; ,
$$
$$
\left ( - { 2 \over   \cos^{2} z }  \;  \hat{a}_{+} \hat{b}
+
 {\partial^{2} \over  \partial z^{2}}
  +2\epsilon M \right )  G_{3}   +2{ \sin z \over \cos^{2} z} \; \hat{a}_{+} \hat{b} \;G_{2} = 0\; ,
$$
$$
\left ( -  {1  \over  \cos^{2} z} (\hat{b}_{-} \hat{a}  + \hat{a}_{+} \hat{b} +2 )
 + {\partial^{2} \over  \partial z ^{2}}  +2\epsilon M   + 1
  \right )  G_{2}
  $$
  $$
  +
 2 {\sin  z \over \cos^{2} z} \; (G_{1} + G_{3}) =0 \; .
\eqno(3.5b)
$$

Now we should define a factorized form for three functions

$$
G_{1} = Z_{1}(z) R_{1}(r)\;, \;
G_{2} = Z_{2}(z) R_{2}(r)\;, \;
G_{3} = Z_{3}(z) R_{3}(r)\; ;
\eqno(3.6a)
$$

\noindent then eqs.  (3.5b) read

$$
\left ( -{ 2\over  \cos^{2} z} \;  \hat{b}_{-} \hat{a}    +
 {\partial^{2} \over  \partial z^{2}}    + 2\epsilon M  \right ) Z_{1} R_{1}  +
    2{ \sin z \over  \cos^{2} z}   \; \hat{b}_{-} \hat{a} \; Z_{2} R_{2}   = 0 \; ,
$$
$$
\left ( - { 2 \over   \cos^{2} z }  \;  \hat{a}_{+} \hat{b}
+
 {\partial^{2} \over  \partial z^{2}}
  +2\epsilon M \right )  Z_{3} R_{3}   +2{ \sin z \over \cos^{2} z} \; \hat{a}_{+} \hat{b} \;Z_{2} R_{2} = 0 \; ,
$$
$$
\left ( -  {1  \over  \cos^{2} z} (\hat{b}_{-} \hat{a}  + \hat{a}_{+} \hat{b} +2 )
 + {\partial^{2} \over  \partial z ^{2}} +2\epsilon M  +1
  \right ) \; Z_{2} R_{2}
  $$
  $$
  +
 2 {\sin  z \over \cos^{2} z} \; (   Z_{1} R_{1} + Z_{3} R_{3} )  =0 \; .
\eqno(3.6b)
$$

Note that the first equation in  (3.6b) does not change if one acts from the left by the operator
 $\hat{b}_{-} \hat{a}$; similarly the second equation preserves its form if on acts from the left by the operator
 $\hat{a}_{+} \hat{b}  $. Therefore, one cam assume
 existence of the following radial relationships
 $$
\hat{b}_{-} \hat{a} R_{1} = \lambda \; R_{1} \;, \qquad \hat{b}_{-} \hat{a} R_{2} = \lambda \; R_{2} \; ,
\qquad R_{1} = R_{2} = R \; ;
\eqno(3.7a)
$$

\noindent
and
$$
\hat{a}_{+} \hat{b} R_{3} = \lambda ' \; R_{3} \;, \qquad \hat{a}_{+} \hat{b} R_{2} = \lambda ' \; R_{2} \;,
 \qquad R_{2}= R_{3}=R \;  .
\eqno(3.7b)
$$

Taking into account these restrictions
from  (3.6b)  we obtain the system in $z$ variable

$$
\left ( -{ 2  \lambda \over  \cos^{2} z} \;    +
 {d^{2} \over  d z^{2}}    + 2\epsilon M  \right ) Z_{1}   +
    2\lambda { \sin z \over  \cos^{2} z}   \;  Z_{2}    = 0 \; ,
$$
$$
\left ( - { 2 \lambda ' \over   \cos^{2} z }
+
 {d^{2} \over  d z^{2}}
  +2\epsilon M \right )  Z_{3}  +2\lambda ' { \sin z \over \cos^{2} z} \; Z_{2}  = 0 \; ,
$$
$$
\left ( -  {1  \over  \cos^{2} z} ( \lambda + \lambda '+2  )
  + {d^{2} \over  d z ^{2}} +2\epsilon M  +1
  \right ) \; Z_{2} +
  2 {\sin  z \over \cos^{2} z} \; (   Z_{1}  + Z_{3}  )  =0 \; .
$$
$$
\eqno(3.8)
$$

With the use of explicit  expressions for operators
 $\hat{a},\hat{a}_{+}, \hat{b},\hat{b}_{-}$,
we derive
$$
\hat{b}_{-} \hat{a}= {1 \over 2} \left (  - {d ^{2} \over d r^{2}} -
 {\cos r \over \sin r}  {d \over d r} - B + {\nu^{2}(r) \over \sin^{2} r} \right )\; ,
 $$
 $$
\hat{a}_{+} \hat{b}= {1 \over 2} \left (  - {d ^{2} \over d r^{2}} -
 {\cos r \over \sin r}  {d \over d r} + B + {\nu^{2}(r) \over \sin^{2} r} \right ) \; ,
 $$
$$
 \hat{a}_{+} \hat{b} =  \hat{b}_{-} \hat{a}    - B \; ,
$$

\noindent
so the first radial equation  for $R_{2}$  takes the form

$$
\hat{b}_{-} \hat{a} \; R_{2} = \lambda \; R_{2} \qquad \Longrightarrow
$$
$$
\left (  { d ^{2} \over d  r^{2}} +
 {\cos r \over \sin r}  { d \over d  r} + B -  {\nu^{2}(r) \over \sin^{2} r} + 2  \lambda  \right ) R_{2} =0 \; ;
\eqno(3.9a)
$$

\noindent
 the second equation for $R_{1}$  gives the same only if two parameters
$\lambda$ and  $\lambda'$ obey a special additional constraint
$$
\hat{a}_{+} \hat{b} \; R_{2} = \lambda ' \; R_{2} \qquad \Longrightarrow \qquad
 \hat{b}_{-} \hat{a}     \; R_{2} = (\lambda '  +B )\; R_{2} ,
$$

\noindent
that is
$$
\lambda' = \lambda - B \; .
\eqno(3.9b)
$$

Let us consider eq.  (3.9a) in more detail

$$
{d^{2} \over dr^{2}} R + {1 \over \mbox{tan}\; r } {d R \over d r }
  -
 {1 \over \mbox{sin}^{2} r   }    [\;  m  +   B  (1 -  \mbox{cos} \;r )] ^{2}\; R   + (B+2 \lambda) \;   R = 0\,.
$$

\noindent
In a new variable
$$
1 - \mbox{cos} \;r  = 2y \; , \qquad  \qquad
 y =  \mbox{sin}^{2} {r \over 2} \in [0 , \; 1 \; ] \,,
$$
$$
\left [ y (1-y) {d^{2} \over dy^{2}} +  (1-2y)  {d \over dy} - \right.
$$
$$
\left. -
 {1 \over 4} ( {m^{2}\over  y} -4B^{2} + {(m+2B)^{2}\over 1-y} \; )\;
+(B+2 \lambda)\right  ] R = 0 \; .
\eqno(3.10)
$$

With the  substitution $
R = y^{a} (1-y)^{b} \; F$,
eq.  $(3.10)$ gives
$$
y (1-y) \; F'' +
[\;  a (1-y)  - b y   + a  (1-y)   - b y  +  (1-2y)  ] \; F'
$$
$$
+ {1 \over y}\;  [ \; a (a-1)  + a  -  {m^{2} \over 4 } \;] \;  F \;\; + \;\;
 {1 \over 1-y }\; [\;   b (b-1)  + b  - {(m+2B)^{2} \over 4 }  \;]\;  F
$$
$$
- [  \; a(a+1)  + 2a b         +   b (b+1)
  - B^{2}     - (B+2 \lambda) \; ] \;  F = 0 \; .
  $$

\noindent
If parameters  obey restriction below
$$
a = \pm {\mid  m  \mid   \over 2 } \;  , \qquad b = \pm { \mid m+2B  \mid  \over 2} \; ;
\eqno(3.11a)
$$

\noindent
we arrive at a more simple equation

$$
 y (1-y) \; F'' +
[\; (2a+1) -2 (a+b+1)y\;]  \; F'
$$
$$
- [  a(a+1)  + 2a b         +   b (b+1)
  - B^{2}     - (B+2 \lambda)  ] \;  F = 0 \; ,
$$
$$
\eqno(3.11b)
$$

\noindent
which is recognized as  a hypergeometric one
$$
y(1-y)\;  F + [ \; \gamma - (\alpha + \beta +1) y \; ]\;  F' - \alpha \beta \; F = 0 \; .
\eqno(3.11c)
$$

So we have (to obtain solutions for bound states we must assume positive $a$  and $b $)
$$
y =  \mbox{sin}^{2}{r \over 2} \; , \qquad  y\, \in \,[\,0 , \, + 1 \,] \,,
\qquad r \in [ \, 0, + \pi  \, ] \,,
$$
$$
R = ( \mbox{sin}\; {r \over 2})^{+\mid  m  \mid } \;\; (\mbox{cos}\; { r \over 2} )^{+ \mid m +2B \mid}
\;\; F (\alpha , \beta, \gamma ; \;  - \mbox{sin}^{2}{r \over 2} ) \,;
$$
$$
\eqno(3.11d)
$$

\noindent
parameters  $(\alpha , \beta, \gamma)$are determined by
$$
\gamma = + \mid m \mid +1 \,, \qquad a = + {\mid m \mid \over 2 }\; , \qquad
b =  + { \mid m+2B  \mid  \over 2} \; ,
$$
$$
\left \{ \begin{array}{l}
 \alpha + \beta  = 2a + 2b +1  \; , \\
\alpha \; \beta  =  (a+b)(a+b+1) - B^{2}     - (B+2 \lambda)  \; ;
\end{array} \right.
 $$
 $$
 \eqno(3.12a)
   $$

\noindent
that is
$$
\gamma =  + \mid m \mid +1 \; , \; a = + {\mid m \mid \over 2 }\; , \;
b =  + { \mid m+2B  \mid  \over 2} \; ,
$$
$$
\alpha = a + b +{1\over 2} - \sqrt{\left(B+{1\over 2}\right)^{2}  +2 \lambda} \; ,
$$
$$
\beta = a + b +{1\over 2} + \sqrt{\left(B+{1\over 2}\right)^{2}  +2 \lambda} \; .
\eqno(3.12b)
 $$

To obtain solutions in polynomials,
we must assume positiveness of the expression under the sign of square root
and must impose restriction on the $\alpha$
$$
\qquad \qquad \qquad \alpha = a + b +{1\over 2} - \sqrt{\left(B+{1\over 2}\right)^{2}  +2 \lambda  } = -n  = 0, -1, -2, ...,
\eqno(3.13a)
$$

\noindent
from whence it follows
the quantization rule
$$
2\lambda    +  \left(B+{1\over 2}\right)^{2} =  (a+b +{1\over 2} + n)^{2} > 0 \; ,
\eqno(3.13b)
$$

\noindent
solutions corresponding to bound states are given by
$$
R = ( \mbox{sin}\; {r \over 2})^{+\mid  m  \mid } \;\; (\mbox{cos}\; { r \over 2} )^{+ \mid m +2B \mid}
$$
$$
\times \; F (-n\;  ,  \mid  m  \mid +  \mid m +2B \mid +1  + n \; , \mid m \mid +1
; \;  - \mbox{sin}^{2}{r \over 2} ) \,.
$$
$$
\eqno(3.13c)
$$

Below, we will use notation
$$
\lambda = \Lambda - {B \over 2}
\eqno(3.14a)
$$

\noindent then the  formula for spectrum (3.13b)will read
$$
2\Lambda    +  B^{2}  =   N(N+1) \; , \qquad N = a+b  + n    \; .
\eqno(3.13b)
$$

\subsection*{4.  Behavior of solutions in $z$ variable  near  singular points
}

Let us turn to the system  (3.8)
$$
\left (  {d^{2} \over  d z^{2}}  -{ 2  \lambda \over  \cos^{2} z} \;        + 2\epsilon M  \right ) Z_{1}   +
    2\lambda { \sin z \over  \cos^{2} z}   \;  \bar{Z}_{2}    = 0 \; ,
$$
$$
\left (  {d^{2} \over  d z^{2}}
 - { 2 \lambda'   \over   \cos^{2} z }
+  2\epsilon M \right )  Z_{3}  +2\lambda '  { \sin z \over \cos^{2} z} \; \bar{Z}_{2}  = 0 \; ,
$$
$$
\left ( {d^{2} \over  d z ^{2}}  -  {\lambda + \lambda'  +2     \over  \cos^{2} z}
     +2\epsilon M  +1   \right )  \bar{Z}_{2}
     +
  2 {\sin  z \over \cos^{2} z} \; (   Z_{1}  + Z_{3}  )  =0 \; .
$$
$$
\eqno(4.1b)
$$

In the variable
$$
\sin z = x \; , \qquad x \in [-1, +1 ]\;,
$$

\noindent
we get
$$
\left (  (1 -x^{2}) {d ^{2} \over  d x^{2}}  - x {d \over dx} - { 2  \lambda \over  1 - x^{2} } \;
    + 2\epsilon M  \right ) Z_{1}   +
     { 2\lambda x  \over  1 - x^{2} }   \;  \bar{Z}_{2}    = 0 \; ,
$$
$$
\left (  (1 -x^{2}) {d ^{2} \over  d x^{2}}  - x {d \over dx} - { 2  \lambda '\over  1 - x^{2} } \;
    + 2\epsilon M  \right ) Z_{3}   +
    {  2\lambda'  x  \over  1 - x^{2} }   \;  \bar{Z}_{2}    = 0 \; ,
$$
$$
\left (  (1 -x^{2}) {d ^{2} \over  d x^{2}}  - x {d \over dx} - { 2+   \lambda  + \lambda '\over  1 - x^{2} } \;
    + 2\epsilon M   + 1 \right ) \bar{Z}_{2}     +
     { 2x  \over  1 - x^{2} }   \;( Z_{1} + Z_{3})    = 0 \; .
$$
$$
\eqno(4.2)
$$

Near the point $z=+ \pi / 2$ we have
$$
z=+ \pi / 2\;, \qquad x \rightarrow +1 \; ,
$$
$$
\left (  2(1-x)  {d ^{2} \over  d x^{2}}  -  {d \over dx} -  {  \lambda \over 1-x}   \;
      \right ) Z_{1}   +
      {\lambda  \over 1-x}    \;  \bar{Z}_{2}    = 0 \; ,
$$
$$
\left ( 2 (1-x)  {d ^{2} \over  d x^{2}}  -  {d \over dx} -  {  \lambda '  \over 1-x }\;
     \right ) Z_{3}   +
      {\lambda'  \over 1-x}      \;  \bar{Z}_{2}    = 0 \; ,
$$
$$
\left (  2 (1-x)  {d ^{2} \over  d x^{2}}  -  {d \over dx} - { 2+   \lambda  + \lambda ' \over 1-x}
    \right ) \bar{Z}_{2}     +
      {1 \over 1-x }    \;( Z_{1} + Z_{3})    = 0 \; ;
$$
$$
\eqno(4.3a)
$$

\noindent so the possible solution is
$$
Z_{1} = A_{1} (1-x)^{a} \;, \qquad
\bar{Z}_{2} = A_{2} (1-x)^{a} \;, \qquad
Z_{3} = A_{3} (1-x)^{a} \; .
\eqno(4.3b)
$$

\noindent
 Substituting (4.3b)  into (4.3a), we obtain
 linear system with respect to
  $A_{1},A_{2},A_{3}$:
$$
(2a^{2}  - a - \lambda )A_{1} + \lambda A_{2} =0 \; ,
$$
$$
(2a^{2}  - a - \lambda' )A_{3} + \lambda' A_{2} =0 \; ,
$$
$$
( 2a^{2}  - a - {2 + \lambda + \lambda ' \over 2} ) A_{2} + A_{1} + A_{3} =0 \; .
\eqno(4.3c)
$$

In similar manner consider behavior od solution near the second singular ponit
$$
z=- \pi / 2\;, \qquad x \rightarrow -1 \; ,
$$
$$
\left (  2(1+x)  {d ^{2} \over  d x^{2}}  +  {d \over dx} -  {  \lambda \over 1+x}   \;
      \right ) Z_{1}   -
      {\lambda  \over 1+x}    \;  \bar{Z}_{2}    = 0 \; ,
$$
$$
\left ( 2 (1+x)  {d ^{2} \over  d x^{2}}  + {d \over dx} -  {  \lambda '  \over 1+x }\;
     \right ) Z_{3}   -
      {\lambda'  \over 1+x}      \;  \bar{Z}_{2}    = 0 \; ,
$$
$$
\left (  2 (1+x)  {d ^{2} \over  d x^{2}}  +  {d \over dx} - { 2+   \lambda  + \lambda ' \over 1+x}
    \right ) \bar{Z}_{2}     -
      {1 \over 1+x }    \;( Z_{1} + Z_{3})    = 0 \; ;
$$
$$
\eqno(4.4a)
$$

\noindent that is
$$
Z_{1} =B_{1} (1+x)^{b} \;, \qquad
\bar{Z}_{2} = B_{2} (1+x)^{b} \;, \qquad
Z_{3} = B_{3} (1+x)^{b} \; .
\eqno(4.4b)
$$

\noindent
and coefficients  $B_{1},B_{2},B_{3}$ obey the linear system as well
$$
(2b^{2}  - b - \lambda )B_{1} - \lambda B_{2} =0 \; ,
$$
$$
(2b^{2}  - b - \lambda' )B_{3} - \lambda' B_{2} =0 \; ,
$$
$$
( 2b^{2}  - b - {2 + \lambda + \lambda ' \over 2} ) B_{2} - B_{1} - B_{3} =0 \; ,
\eqno(4.4c)
$$

With the notation
$$
2a ^{2}  - a  = A\;, \qquad 2b^{2}  - b = B \; ,
$$
$$
a= {1 \pm \sqrt{1+ 8A} \over 4}\;, \qquad  b= {1 \pm \sqrt{1+ 8B}\over 4}\; ;
\eqno(4.5a)
$$

\noindent
two linear system are written as
$$
(A - \lambda )A_{1} + \lambda A_{2} =0 \; ,
$$
$$
(A - \lambda' )A_{3} + \lambda' A_{2} =0 \; ,
$$
$$
( A - {2 + \lambda + \lambda ' \over 2} ) A_{2} + A_{1} + A_{3} =0 \; ;
\eqno(4.5b)
$$

\noindent and
$$
(B - \lambda )B_{1} - \lambda B_{2} =0 \; ,
$$
$$
(B - \lambda' )B_{3} - \lambda' B_{2} =0 \; ,
$$
$$
( B- {2 + \lambda + \lambda ' \over 2} ) B_{2} - B_{1} - B_{3} =0 \; ,
\eqno(4.5c)
$$

Further we get one the same eigenvalue equation for values
$A$ and   $B$
$$
(A- \lambda ) \lambda ' + (A- \lambda') \lambda  -(A- \lambda) (A - \lambda') (A -
{2+ \lambda + \lambda ' \over 2})=0 \; ,
$$
$$
(B- \lambda ) \lambda ' + (B- \lambda') \lambda  -(B- \lambda) (A -
 \lambda')  (B- {2+ \lambda + \lambda'  \over 2})=0 \; ;
\eqno(4.6)
$$

\noindent
 respective solutions are given as
$$
A_{1} = (A_{2} ) \; { \lambda  \over  \lambda - A } \; , \qquad
A_{3} = (A_{2} ) \; { \lambda'  \over  \lambda' - A } \; ;
\eqno(4.7a)
$$
$$
B_{1} = ( -B_{2} ) \; { \lambda  \over  \lambda - B } \; , \qquad
B_{3} = (- B_{2} ) \; { \lambda'  \over  \lambda' - B } \; .
\eqno(4.7b)
$$

Now, let us examine a third order equation  (4.6) --
for definiteness consider the case of $A$:
$$
2(A- \lambda ) \lambda ' + 2(A- \lambda') \lambda  -  (A- \lambda) (A - \lambda')
(2 A - 2 - \lambda - \lambda'  )=0 \; ;
$$
$$
\eqno(4.8)
$$

\noindent the equation arising is symmetric with respect
to formal replacement
$\lambda \Leftrightarrow \lambda '$.
Explicitly the equation read
$$
2A (\lambda  + \lambda' ) - 4\lambda \lambda' +
[A^{2}  - A (\lambda  + \lambda' ) + \lambda \lambda ' ] [ -2 A + 2 + (\lambda + \lambda' ) ] = 0 \qquad \Longrightarrow
$$

$$
2A (\lambda  + \lambda' ) - 4\lambda \lambda'   -2 A^{3}  + 2 A^{2} + A^{2} (\lambda + \lambda' )
$$
$$
+ 2 A^{2} (\lambda + \lambda' )  - 2A (\lambda + \lambda' )  - A(\lambda + \lambda' )^{2}
-2 A \lambda \lambda'  + 2  \lambda \lambda' + \lambda \lambda' (\lambda + \lambda' ) = 0
\qquad \Longrightarrow
$$

$$
-2 A^{3} + A^{2} \; [ \;  2 + 3 (\lambda + \lambda') \; ]   - A \; [\;    (\lambda + \lambda' )^{2}
+2  \lambda \lambda' \;  ] +  \lambda \lambda ' \; [\; ( \lambda + \lambda')   -2   \; ]  = 0\; .
\eqno(4.9a)
$$

Remembering on  $\lambda ' = \lambda -B$, one can introduce other parameters
$$
\lambda ' -{B \over 2} =  \lambda  +{B \over 2} \equiv \Lambda \; ,
$$
$$
\lambda + \lambda' = 2 \Lambda\;, \qquad \lambda  \lambda' = \Lambda^{2} - {B^{2} \over 4} \; .
\eqno(4.9b)
$$

\noindent Then eq. (4.9a) reads
$$
 A^{3} - A^{2} \; (   3 \Lambda +1 )  + A \; (   3 \Lambda ^{2}  - {B^{2} \over 4} ) -
   ( \Lambda^{2} - {B^{2} \over 4}) \; (   \Lambda  -1 )   = 0\; .
\eqno(4.9c)
$$

It can be  presented symbolically as

$$
A^{3} + a A^{2} + b A + c = 0 \; ,
\eqno(4.10a)
$$

\noindent
where
$$
a = - (   3 \Lambda +1 )\;,
$$
$$
b = (   3 \Lambda ^{2}  - {B^{2} \over 4} )\; ,
$$
$$
c= - ( \Lambda^{2} - {B^{2} \over 4}) \; (   \Lambda  -1 )\; .
\eqno(4.10b)
$$

Through  change in the  variable
 ($A \Longrightarrow Y$)
$$
A = Y -{a \over 3} = Y + \Lambda + {1 \over 3}
\eqno(4.11a)
$$

\noindent
we remove a quadratic term

$$
Y^{3} + p Y + q =0 \; ,
\eqno(4.11b)
$$

\noindent where
$$
p =  -{a^{2} \over 3} + b = -( 2\Lambda + {B^{2} \over 4} + { 1 \over 3})   \;,
$$
$$
q = {2a^{3} \over 27} - {ab \over 3} + c = - ( {2 \over 3} \Lambda + {B^{2} \over 3} + {2 \over 27} )  \; .
\eqno(4.11c)
$$

\noindent Note substantial inequalities

$$
p < 0, \qquad q < 0, \qquad \mid p \mid > \mid q \mid \; .
$$

Formulas, giving solutions of eq.  (4.11b) are well known

$$
Y = \left [ -{q \over 2} + \sqrt{({q \over 2})^{2}  + ({p \over 3})^{3}}\;  \right ]^{1/3} +
\left [ -{q \over 2} - \sqrt{({q \over 2})^{2}  + ({p \over 3})^{3}}\;  \right ]^{1/3} \; .
\eqno(4.12a)
$$

Applying (4.12a), one must use correlated roots

$$
\alpha = \left [ -{q \over 2} + \sqrt{({q \over 2})^{2}  + ({p \over 3})^{3}}\;  \right ]^{1/3}
\eqno(4.12b)
$$

\noindent
and

$$
\beta = \left [ -{q \over 2} - \sqrt{({q \over 2})^{2}  + ({p \over 3})^{3}}\;  \right ]^{1/3}
\eqno(4.12c)
$$

\noindent
so that the following restriction hold
$$
\alpha \beta = -{p \over 3}\; .
\eqno(4.12d)
$$

Besides, the roots can be searched according to the formulas

$$
Y_{1} = \alpha _{1} + \beta_{1} \;,
$$
$$
Y_{2} = -{1 \over 2} ( \alpha _{1} + \beta_{1} ) + i {\sqrt{3}  \over 2} (\alpha _{1} - \beta_{1})
$$
$$
Y_{3} = -{1 \over 2} ( \alpha _{1} + \beta_{1} ) - i {\sqrt{3}  \over 2} (\alpha _{1} - \beta_{1})
\eqno(4.13a)
$$

\noindent
where  $\alpha_{1}$  stands for any root in (4.12b), but a root $\beta_{1}$ in (4.12c) must obey

$$
\alpha _{1} \beta_{1}  = -{p \over 3} \; .
\eqno(4.13b)
$$

Let us additionally detail expressions (4.13a,b)  for three  roots. Allowing for

$$
\alpha = \left [ - q / 2  +  i \sqrt{    (-p / 3)^{3} - ( q / 2)^{2} }\;  \right ]^{1/3}
$$
$$
=
\left [  (-p / 3)^{3/2} (\cos \phi + i \sin  \phi )  \right ]^{1/3}
$$
$$
=
\sqrt{-p/3}\;  \;
\left \{   e^{i \phi/3 }  \; , \; e^{i ( \phi/3  +2\pi /3)}\; , \;  e^{i ( \phi/3  +4\pi /3)} \; \right  \} \; ,
\eqno(4.14a)
$$

\noindent
where

$$
\cos \phi = { -q / 2 \over (-p / 3)^{3/2} } \;, \qquad
\sin \phi = { \sqrt{  (-p / 3)^{3} - ( q / 2)^{2}  }  \over (-p / 3)^{3/2} } \; .
\eqno(4.14b)
$$

It is readily to specify the quantity
 $\beta$:

$$
\beta = \left [ - q / 2  -  i \sqrt{    (-p / 3)^{3} - ( q / 2)^{2} }\;  \right ]^{1/3}=
$$
$$
=
\left [  (-p / 3)^{3/2} (\cos \phi - i \sin  \phi )  \right ]^{1/3} =
$$
$$
=
\sqrt{-p/3}\;  \;
\left \{   e^{-i \phi/3 }  \; , \; e^{i ( - \phi/3  +2\pi /3)}\; , \;  e^{i ( - \phi/3  +4\pi /3)} \; \right  \} \; ,
\eqno(4.14a)
$$

\noindent
where

$$
\cos \phi = { -q / 2 \over (-p / 3)^{3/2} } \;, \qquad
\sin \phi = { \sqrt{  (-p / 3)^{3} - ( q / 2)^{2}  }  \over (-p / 3)^{3/2} } \; .
\eqno(4.14b)
$$

As  $\alpha_{1}$ and   $\beta_{1}$  we will take

$$
\alpha_{1} =  \sqrt{-p/3} e^{+i\phi /3} \;, \qquad
\beta_{1} =  \sqrt{-p/3} e^{-i\phi /3} \; ;
$$
$$
\cos \phi = { -q / 2 \over (-p / 3)^{3/2} } \;, \qquad
\sin \phi = { \sqrt{  (-p / 3)^{3} - ( q / 2)^{2}  }  \over (-p / 3)^{3/2} } \; .
\eqno(4.15a)
$$

And further we readily find

$$
\alpha _{1} + \beta_{1} = 2 \sqrt{-p/3} \; \cos {\phi \over 3} \;, \qquad
\alpha _{1} - \beta_{1} = 2 i \sqrt{-p/3} \; \sin {\phi \over 3}\; .
\eqno(4.15b)
$$

Thus, three different (real-valued) roots are determined by the formulas

$$
Y_{1} =  \sqrt{-p/3} \; (2 \cos {\phi \over 3}  ) \;,
$$
$$
Y_{2} = \sqrt{-p/3} \; (  -  \cos {\phi \over 3}  -  \sqrt{3}    \sin {\phi \over 3}) \; ,
$$
$$
Y_{3} = \sqrt{-p/3} \; ( -  \cos {\phi \over 3} + \sqrt{3}   \sin {\phi \over 3} ) \; .
\eqno(4.16)
$$

One can additionally  check the results: from
 the identity

$$
Y^{3} + p Y + q = (Y-Y_{1})(Y-Y_{2})(Y-Y_{3})
$$
it follows
$$
0 = Y_{1} + Y_{2} + Y_{3} \; ,
$$
$$
p = Y_{1}Y_{2} +  Y_{1}Y_{3}  + Y_{2}Y_{3}  \; , \qquad
q = - Y_{1}Y_{2} Y_{3} \; .
\eqno(4.17)
$$

First we readily verify two identity
$$
0 = Y_{1} + Y_{2} + Y_{3} \; , \qquad
p = Y_{1}Y_{2} +  Y_{1}Y_{3}  + Y_{2}Y_{3}  \; .
$$

\noindent   Turning to the third ine, leyt us calculate
$$
- Y_{1}Y_{2} Y_{3} = -
{2 \sqrt {3}\over 9}\,(-p)^{3/2}\,
\left[4\,\cos^{2}{\phi\over 3}-3\right]\, \cos {\phi\over 3}\,;
\eqno(4.18a)
$$

\noindent further with the help of elementary relation
$$
\cos \alpha \cos \beta = { \cos (\alpha -\beta) + \cos (\alpha + \beta) \over 2} \; ,
$$

\noindent
we get
$$
\left[4\,\cos^{2}{\phi\over 3}-3\right]\, \cos {\phi\over 3} = ( -1 + 2 \cos {2\phi \over 3} )  \cos {\phi \over 3}  = \cos \phi \;;
\eqno(4.18b)
$$

\noindent
and thus we prove the third identity (remembering on  (4.15a))
$$
- Y_{1}Y_{2} Y_{3} = -
{2 \sqrt {3}\over 9}\,(-p)^{3/2}  \cos \phi = {2 \sqrt {3}\over 9}\,(-p)^{3/2} \;
{ -q / 2 \over (-p / 3)^{3/2} } = - q \; .
\eqno(4.18c)
$$

Unfortunately we have not gained success  in solving the  main system of 3 equation in $z$ variable
$$
\left (  (1 -x^{2}) {d ^{2} \over  d x^{2}}  - x {d \over dx} - { 2  \lambda \over  1 - x^{2} } \;
    + 2\epsilon M  \right ) Z_{1}   +
     { 2\lambda x  \over  1 - x^{2} }   \;  \bar{Z}_{2}    = 0 \; ,
$$
$$
\left (  (1 -x^{2}) {d ^{2} \over  d x^{2}}  - x {d \over dx} - { 2  \lambda '\over  1 - x^{2} } \;
    + 2\epsilon M  \right ) Z_{3}   +
    {  2\lambda'  x  \over  1 - x^{2} }   \;  \bar{Z}_{2}    = 0 \; ,
$$
$$
\left (  (1 -x^{2}) {d ^{2} \over  d x^{2}}  - x {d \over dx} - { 2+   \lambda  + \lambda '\over  1 - x^{2} } \;
    + 2\epsilon M   + 1 \right ) \bar{Z}_{2}     +
     { 2x  \over  1 - x^{2} }   \;( Z_{1} + Z_{3})    = 0 \; .
$$

So this analysis canon be considered as completed.

\subsection*{ Acknowledgment }

Authors are grateful to Professor A.L. Sanin and Professor  D.V. Serov
for warm welcome in Saint  Petersburg State Polytechnical University, and
for friendly  encouragement  and helpful  advices.

\end{document}